\documentstyle[aps,prl,multicol,psfig]{revtex}

\begin{document}


\title{Forest-fire models as a bridge between different paradigms in Self-Organized
Criticality}

\author{Proshun Sinha-Ray and Henrik Jeldtoft Jensen }

\address{Department of Mathematics, Imperial College,
180 Queens Gate, London SW7 2BZ, United Kingdom}

\twocolumn[\hsize\textwidth\columnwidth\hsize\csname %
@twocolumnfalse\endcsname

\maketitle
\begin{abstract}
We turn the stochastic critical forest-fire model introduced by Dr\"ossel and Schwabl (PRL {\bf 69},
1629, 1992) into a deterministic threshold model. This new model has many features in 
common with sandpile and earthquake models of Self-Organized Criticality. Nevertheless our 
deterministic forest-fire model exhibits in detail the same macroscopic statistical properties as 
the original Drossel and Schwabl model. We use the deterministic model and a related 
semi-deterministic version of the model to elaborate on the relation between forest-fire, 
sandpile and earthquake models.   
\end{abstract}


]

\narrowtext

Several types of models of self-organised criticality (SOC) exist\cite{Bak:book,Jensen:book}.
The original cellular automaton models were defined by a deterministic and conservative 
updating algorithm, with thresholds (barriers to activity), and stochastic driving
\cite{BTW:SOC1,BTW:SOC}. A new variation of models was developed by Olami, Feder and Christensen 
(OFC) \cite{OFCmodel} who realised that a non-conservative threshold model might remain critical 
if driven uniformly. The OFC model is completely deterministic except for a random initial 
configuration. In both types of model the threshold is assumed to play a crucial role as 
a local rigidity which allows for a separation of time scales and, equally important, 
produces a large number of metastable states. The dynamics take the system from one of 
these metastable states to another. It is believed that separation of time scale and 
metastability are essential for the existence of scale invariance in these models.

A seemingly very different type of model was developed by Dr\"ossel and Schwabl 
(DS)\cite{DroSchff}. No threshold appears explicitly in this model and the separation of 
time scales is put in by hand by tuning the rates of two stochastic processes which 
act as driving forces for the model. The DS forest-fire (FF) is defined on a $d$-dimensional 
square lattice. Empty sites are turned into ``trees'' with a probability $p$ per site in 
every time step. A tree can catch fire stochastically when hit by ``lightning'', with 
probability $f$ each time step, or deterministically when a neighbouring site is on fire. 
The model is found to be critical in the limit $p\rightarrow0$ together with 
$f/p\rightarrow0$. This model is a generalization of a model first suggested by Bak, Chen 
and Tang \cite{BCTff} which is identical to the DS model except that it does not contain 
the stochastic ignition by lightning. The BCT system is not critical \cite{BCTffnotSOC}
(in less than three dimensions, see \cite{BCTffin3d}). A continuous variable, uniformly 
driven deterministic version \cite{detff} also shows regular behavior for low values 
of $p$\cite{detffnotSOC}. Thus the introduction of the stochastic lightning mechanism 
appeared to be necessary, at least in two dimensions, for the model to behave critically. 
A useful review can be found in \cite{ffrev}.

In the present letter we present a transformation of the forest-fire model into a completely 
deterministic system. This model is an extension of the recently introduced auto-ignition 
forest-fire, a simple variation on the DS model\cite{autoigff}. As in that model, we find that all 
macroscopic statistical measures of the system are preserved. Specifically, we show that the three 
models have the same exponent for the probability density describing clusters of trees, similar 
probability densities of tree ages and, probably most unexpected, almost 
the same power spectrum for the number of trees on the lattice as a function 
of time. It is surprising that the temporal fluctuation spectrum can be the same in the 
deterministic model as in the DS forest fire, since even a small stochastic element in an 
updating algorithm is known to be capable of altering the power spectrum in a significant 
way \cite{jens-anders-grin}.

{\it Definition of model -- } The SOC FF can be recast into an auto-ignition model. 
This model is identical to the DS model, except that the spontaneous ignition probability $f$ 
is replaced by an auto-ignition mechanism by which trees ignite automatically 
when their age $T$ after inception reaches a value $T_{max}$. Choosing this value suitably 
with respect to $p$ gives a system with exactly the same behaviour and statistical properties 
as the DS model\cite{autoigff}. Thus one stochastic driving process has been removed and a 
threshold introduced, while maintaining the SOC state; this model also displays explicitly 
the relationship between threshold dynamics and the separation of time scales so necessary 
for the SOC state.

The auto-ignition model can be turned into a completely deterministic critical model
by eliminating  the stochastic growth mechanism. In the deterministic model (which we shall 
call the regen FF) each cell is given an integer parameter $T$ which increases by one each 
time step. If $T>0$, the cell is said to be occupied, otherwise it is empty 
(or regenerating). The initial configuration is a random distribution of $T$-values 
and fires. Fires spread through nearest neighbours and the auto-ignition mechanism 
is again operative so that a tree catches fire when its $T=T_{max}$. However, in 
this model when a tree catches fire the result is a decrement of $T_{regen}$ from 
its $T$-value. Note that when $T_{regen}<T_{max}$, a cell may still
be  occupied 
after it has been ignited. The parameters $T_{max}$ and $T_{regen}$ can be 
thought of as having a qualitatively reciprocal relationship with $f$ and $p$ 
respectively (in terms of the average `waiting time' for spontaneous ignition 
and tree regrowth), though this is less straightforward in the latter case 
because  trees are not always burned down by fire. It is evident that $T_{regen}$ 
also sets, and allows direct control of, the degree of dissipation of the 
$T$-parameter in the system. 

{\it Results -- } 
We now turn to a comparison between the statistical properties of the stochastic DS FF 
and the entirely deterministic regen model, with reference to the partly deterministic 
auto-ignition model. 

\begin{figure}
\psfig{figure=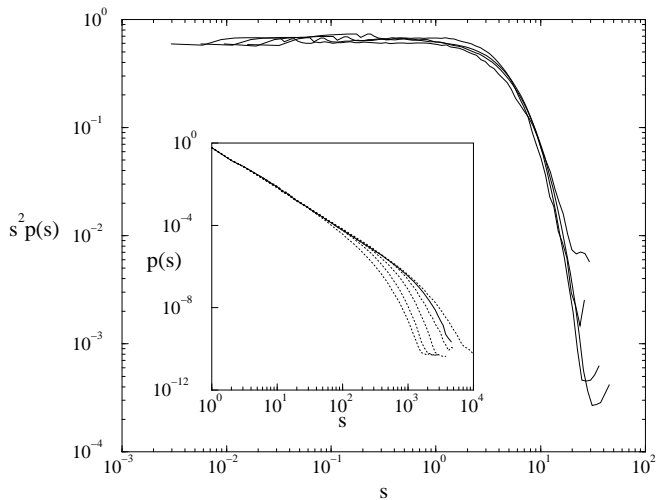,width=\columnwidth}
\caption{\label{regenCl} Scaling plots of tree cluster distributions for regen model ($L=1000$): 
			 $T_{max}=200$ and $t=T_{max}/T_{regen}=0.875, 1.0$; $T_{max}=1000$ 
			 and $t=1.1$. Inset: Scaling of cut-off with $t$ increasing left to right
			 (dotted) and sample distribution for DS model with $p=0.001, f/p=0.01$ 
			 for comparison (solid). }
\end{figure}

First we consider the probability density $p(s)$ of the tree clusters sizes \cite{cluster} 
simulated for different parameters for the different models. It is well 
known that the correlation length in the DS model (as measured by the cut-off $s_c$ in 
$p(s)$) increases as the critical point is approached by decreasing $p$, $f$ and $f/p$ 
\cite{DroSchff}. There is a corresponding increase in the power law regime for the 
cluster distribution in the auto-ignition model as $p$ is decreased and $T_{max}$ 
is increased\cite{autoigff}. 
The scaling behaviour of the cut-off $s_c$ is difficult to 
ascertain due to the limited range of data available, 
but seems to be of the form $\ln(S_c)\sim pT_{max}$, although we cannot 
exclude an algebraic dependence of the form
$s_c\sim 
(pT_{max})^a$, with $a\simeq 6$.
Fig. \ref{regenCl}
 shows scaling plots for the regen model, and we see that here too
the cut-off $s_c$ scales with increasing ratio, $t=T_{max}/T_{regen}$. 
We have approximately $\ln(s_c)\sim T_{max}$  though again the relation
may be algebraic. 
The conclusion is that all three 
models approach a critical state described by the {\em same} power law $p(s)\sim s^{-\tau}$ 
with $\tau\simeq 2$. 

\begin{figure}
\psfig{figure=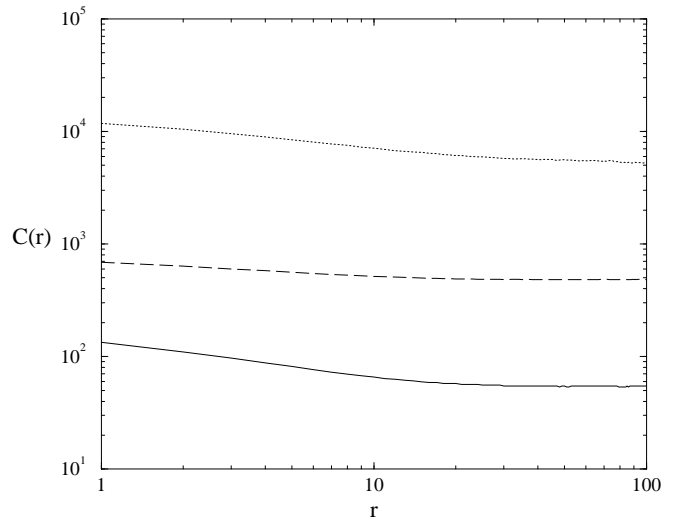,width=\columnwidth}
\caption{\label{T-Tfig} $T-T$ correlation functions for DS (L=1000; $p=0.001, f/p=0.01$ - dotted)
			and regen (L=2000; $T_{max}=200$ and $t=1.0$, trees and empty sites - solid 
			and long dashed) models. Power law correlations are clearly observed. }
\end{figure}

One expects the power law observed in the cluster size distribution to be reflected 
in power laws for spatial correlation functions. It is particularly interesting to 
study the age-age correlation function:
	\begin{equation}
		C(r) = \langle T({\bf r}+{\bf r}_0)T({\bf r}_0)\rangle - \langle T({\bf r_0})\rangle^2
	\label{correlation}
	\end{equation}
This correlation function was never studied for the DS FF because the model does not 
consider the age $T({\bf r})$ explicitly. In Fig. \ref{T-Tfig} we show the behavior of the 
age-age correlation function in the regen and DS models.

As usual it is difficult to obtain a substantial power law region because of finite size 
limitations. Nevertheless it is clear that $C(r)$ does exhibit power law dependence 
on $r$ and we find $C(r)\sim r^{-\eta}$ with  $\eta\simeq 0.32, 0.21$ and $0.23$ for the 
regen, auto-ignition and DS models respectively. 
Interestingly, the same correlation function for empty sites (which have 
negative $T$ in the regen model is also a power law with $\eta\simeq 0.13$. 

\begin{figure}
\psfig{figure=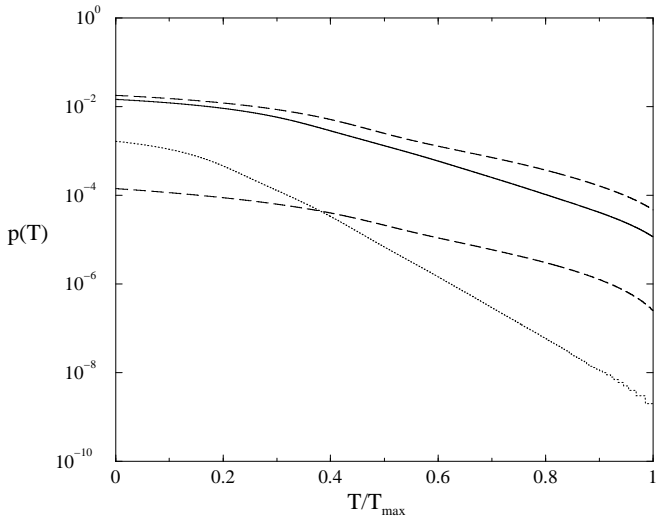,width=\columnwidth}
\caption{\label{agefig} Age profiles for DS ($p=0.001, f/p=0.01$, plotted with a $T_{max}$ of 4000 
			for clarity - dotted), auto-ignition (p=0.0001, $T_{max}=20000$ and 	
			$T_{max}=24000$ - dashed and solid) and regen ($T_{max}=1000$, $t=1.0$ 
			- long dashed) models.}
	                
\end{figure}

Let us now turn to the temporal characteristics of the models. In Fig. \ref{agefig} we show that 
the probability distribution of the ages of the trees has a very 
similar form for all three models. 

All are broad and exponential in character. Since it is a microscopic property, it is 
not surprising that there is some variation between the models. This variation may also be 
the reason for the different exponents in the age-age correlation functions mentioned above.

It is remarkable that the DS FF exhibits a cut-off in the age distribution 
which is nearly as sharp as the cut-off in the two threshold models. This shows that 
the stochastic ignition process in the DS model, characterized by the lightning 
probability $f$, can be replaced in surprising detail by the deterministic age threshold.

\begin{figure}
\psfig{figure=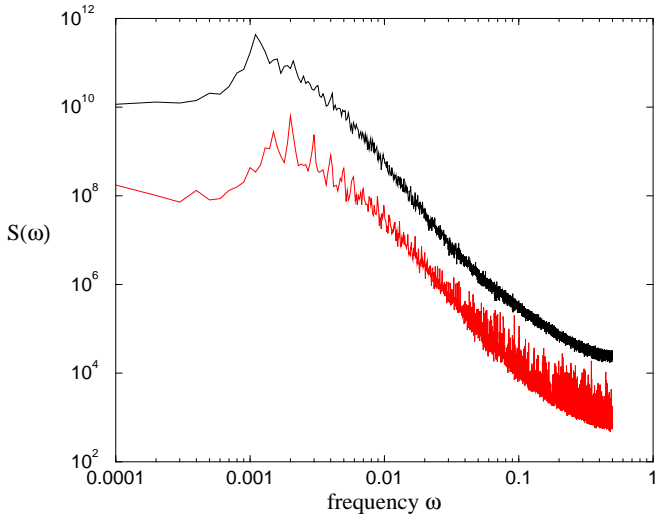,width=\columnwidth}
\caption{\label{nTfig} Tree density power spectra, for DS ($f/p=0.01$ - black) and 
	               regen ($T_{regen}/T_{max}=1.0$ - grey) models}
\end{figure}

The collective temporal behaviour is represented by the power spectrum of the time 
variation of the total number of trees on the lattice. In Fig. \ref{nTfig} these power spectra 
are shown for the DS and regen models (again, the power spectrum for the auto-ignition model is 
nearly identical).

Our most surprising result is that the deterministic regeneration model has nearly the 
same power spectrum as the two other models, particularly in the light of the differences 
in the age profiles above. 

The equivalence between the three models allows us to think of the probabilistic growth 
and lightning in the DS FF model as effectively acting as thresholds. Qualitatively one 
can readily see that the probabilistic nature of the growth and the lightning can be 
interpreted as a kind of rigidity. Namely, an empty site has a rigidity against being 
turned into a tree described by $1/p$. A tree has rigidity against fire described by the 
fact that a tree only catches fire if nearest neighbor to a fire or when hit by lightning.

{\it Discussion -- } 
We now discuss the relationship between the regen model presented above and other SOC models.

Our regen model is similar to the deterministic model introduced by Chen, Bak and 
Jensen\cite{detff}. The crucial difference however, is that in the previous model the 
ratio $T_{regen}/T_{max}$ - which must be decreased to move closer to the critical 
point and obtain scale free behavior - is effectively held fixed at a finite value, 
and hence the model does not allow one to approach the critical state.

The regen model has several features in common with the sandpile and earthquake 
models. It is similar to both sets of models in that the intrinsic dynamics is 
entirely deterministic and controlled by thresholds. The model is uniformly driven like
the OFC earthquake model \cite{OFCmodel} , and moreover, our 
deterministic FF model is genuinely non-conservative. It is worth noting that distributing the 
increase in $T$ randomly in a limited number of portions (rather than equally across all trees) 
each time step was found to destroy the criticality as the size of the portions increased. In one 
important respect our model is more similar to the BTW sandpile model than to the OFC model. 
Namely, when a site suffers relaxation (a tree catches fire) a fixed amount $T_{regen}$ 
is subtracted from the dynamical variable of that site. The same happens in the 
BTW model. In the OFC model, on the other hand, the dynamical variable of a relaxing 
site is reset to zero. This property has been argued to allow for a marginal
synchronisation in the model and hence to be responsible for the OFC model's ability, in 
contrast to the BTW model, to remain critical even in the non-conservative regime\cite{middleton}. 
Seen in this context the deterministic FF model presented here constitutes a very 
interesting mix of features from the BTW and OFC models. Our regen FF model is 
non-conservative, uniformly driven and though the microscopic update does not support a 
marginal synchronisation, nevertheless the model does exhibit the same scale free behavior 
as the DS FF. 

This gives a direct link between the SOC behaviour of the BTW, OFC and DS FF models, each of 
which are commonly assumed to be representative of different and distinct classes of SOC models
(e.g. in \cite{uniMFpic}). Furthermore, the change in the mechanism 
for the renewal of the forest (from a probability for growth to a time for 
regeneration) and the resultant sandpile-like picture allows the identification 
of $p$ with a dissipation parameter (in terms of the subtraction of $T_{regen}$ 
on ignition) rather than as a driving parameter. This is quite contrary to 
the normally held and most obvious view - for the DS FF - that $p$ is the driving parameter 
(creating trees in the system), and that if anything $f$ controls the dissipation (the complete 
combustion of trees into empty sites). If this is so, we can speculate that it may be possible 
to relate the physical limits for critical behaviour in the BTW sandpile: 
	\begin{center}
	$h,h/\epsilon\rightarrow0$ 
	\end{center}
(where $h$ is the driving rate and $\epsilon$ the dissipation) and, recalling the 
qualitatively reciprocal relationships between $f,p, T_{max}$ and 
$T_{regen}$ noted earlier, 
the DS and regen forest-fire models:
	\begin{center}
	$f,f/p\rightarrow0$, and 1/$T_{max}$,$T_{regen}$/$T_{max}\rightarrow0$
	\end{center}

The main difference between the deterministic FF model and the sandpile and earthquake models 
is that the dynamical variable $T$ is {\em not} transported to neighboring sites when a 
site relaxes and that the threshold exists only for the initiation and not propagation of avalanches. 
This difference can be summarized as the FF model being a model of two coupled fields, 
fires and trees, whereas the sandpile and earthquake models contain one self-coupled field, 
the energy of a site. 

Another difference consists in that the thresholds of the deterministic FF model must be 
tuned (to infinity) for the model to approach the critical regime. The reason for this is 
that the thresholds relate directly to the rate of driving in the model. The 
sandpile and earthquake models are different in that the SOC limit of slow driving can be 
reached without a tuning of the thresholds.

Finally, we note that the regen model is critical with periodic boundary conditions (in contrast 
to the BTW and OFC models) and without external driving (unlike the DS model), and is therefore 
the only system which can be said to be completely self-contained.

{\it Conclusion --}  We have demonstrated that the stochastic Drossel-Schwabl forest-fire 
model can be turned into a deterministic threshold model without changing any of the collective
statistical measures of the system in a significant way. The model illuminates greatly the 
relationship between different types of SOC models.

{\it Acknowledgements --}  
HJJ is supported by EPSRC under contract GR/L12325. PSR is the recipient of an EPSRC PhD studentship. 
We would like to thank Barbara Dr\"ossel and Kim Christensen for helpful discussion and insight.


\end{document}